\documentclass[prl,twocolumn,superscriptaddress]{revtex4-2}
\usepackage{graphicx} 
\usepackage{amsmath}
\usepackage{xcolor}

\begin{document}

\title{Fluidization induced by Magnetic Interactions in Confined Active Matter}
\author{Marco Musacchio}
\affiliation{Dipartimento di Fisica,
Sapienza Universit\`a di Roma, Piazzale A. Moro 2, I-00185, Rome, Italy}
\affiliation{Institut f{\"u}r Theoretische Physik II: Weiche Materie,
		Heinrich-Heine-Universit{\"a}t D{\"u}sseldorf, Universit{\"a}tsstra{\ss}e 1,
		D-40225 D{\"u}sseldorf, 
		Germany}

\author{Markus Felber}
\affiliation{Institute of Science and Technology Austria, Am Campus 1, 3400 Klosterneuburg, Austria}

\author{Matteo Paoluzzi}
\affiliation{Dipartimento di Fisica,
Sapienza Universit\`a di Roma, Piazzale A. Moro 2, I-00185, Rome, Italy}

\author{Andrea Gnoli}
\affiliation{
Istituto dei Sistemi Complessi - Consiglio Nazionale delle Ricerche, and Dipartimento di Fisica,
Sapienza Universit\`a di Roma, Piazzale A. Moro 2, I-00185, Rome, Italy}

\author{Andrea Puglisi}
\affiliation{
Istituto dei Sistemi Complessi - Consiglio Nazionale delle Ricerche, and Dipartimento di Fisica,
Sapienza Universit\`a di Roma, Piazzale A. Moro 2, I-00185, Rome, Italy}
\affiliation{INFN, Sezione Roma2, Via della Ricerca Scientifica 1, I-00133, Rome, Italy}

\author{Luca Angelani}
\affiliation{
Istituto dei Sistemi Complessi - Consiglio Nazionale delle Ricerche, and Dipartimento di Fisica,
Sapienza Universit\`a di Roma, Piazzale A. Moro 2, I-00185, Rome, Italy}

\begin{abstract}
We investigate magnetic active matter in confined geometries using both experiments with magnetic toy robots Hexbugs and simulations of elongated magnetic active Brownian particles in circular domains. Standard active particles tend to accumulate at boundaries, forming clusters even at relatively low densities. In the presence of magnetic interactions, we provide evidence for a {\it fluidization} effect that inhibits clustering and shifts its onset to higher packing fractions.
Moreover, magnetic dipolar interactions give rise to novel collective behaviors, such as train-like formations, rotating pairs, and particle vortices.
\end{abstract}

\maketitle


{\em Introduction}. Originally, the passage from chaotic to ordered behavior has been investigated in molecular systems: this is what goes under the name of phase transitions and typically involves very large numbers of particles at thermodynamic equilibrium~\cite{toda1978statistical,ma1985statistical}. 
Since  decades, physicists in the field of complex systems have directed their attention to the emergence of spontaneous order in phenomena not involving molecular matter, typically violating the rules of equilibrium, because of relevant energetic transformations, such as friction and self-propulsion~\cite{van2024soft,sarracino2025nonequilibrium}. Such kinds of systems embrace - usually - ``macroscopic molecules'' such as granular and active particles. The former are passive particles which - as a consequence of their size - interact through inelastic collisions~\cite{brilliantov2010kinetic,puglisi2014transport}. 
The latter are particles propelled by some kind of internal motor that converts stored energy into persistent motion~\cite{bechinger2016active,marchetti2013hydrodynamics}. 
Examples of active particles range 
from micron-size, as in the case of bacteria and self-propelled colloids, 
to the macroscale, as in the case of fishes, birds and robots \cite{ballerini2008interaction, bechinger2016active, scholz2018rotating, workamp2018symmetry, siebers2023exploiting, caprini2024dynamical, volpe2025roadmap, antonov2025self}. 

Among a plethora of collective phenomena emerging in active matter \cite{PhysRevX.12.010501,Elgeti_RPP2015}, two of them are peculiar of the active systems we study here:  accumulation at boundaries and flocking~\cite{PhysRevLett.75.1226}. 
The former is related to the tendency of active particles to accumulate near confining walls, 
due to the persistent character of the motion \cite{Li_2009,Elgeti_2013,Elgeti_2015,Lowen_2008,Angelani_2017,Angelani_2023},
closely related to  motility-induced phase separation (MIPS) \cite{Tailleur08}, which occurs in the bulk.
Flocking is the transition observed when aligning self-propelled particles produce polar order over large scales~\cite{PhysRevLett.75.4326,paoluzzi2024flocking, musacchio2025flocking}. 
Active particles exhibiting magnetic properties are particularly intriguing \cite{klapp2016collective, compagnie2025magnetically, rosenberg2025windmilling}, especially in biological contexts. In magnetotactic bacteria, magnetic interactions not only govern the orientation of individual cells with respect to the Earth's magnetic field, but also influence collective behavior through dipolar interactions between magnetosomes \cite{MagnBact2024}.

Recently, small robots such as Hexbugs nano\textsuperscript{\textregistered} \cite{Hexbug}
and other similar motorized particles have been proposed as a flexible prototype for the experimental study of active matter~\cite{Volpe2024,Dauchot2019,PhysRevResearch.2.043299,Escobar2025,Ning2024,Vanesse2023,Sepulveda2021,Obreque2024, chor2023many_body_szilard, carrillo2025depinning}. 
They are a remarkably rich source of novel phenomena that can be explored in the laboratory without requiring expensive technology or specialized expertise in biological fields.

In~\cite{deblais2018boundaries}, it has been shown that Hexbugs, without any magnetic interaction, can display, above a relatively small packing fraction, a drop-like clustering behavior stabilised by the hard boundary. Here we study the interplay between boundary-induced collective phenomena and magnetic interactions in systems of Hexbugs, conducting experiments and numerical simulations. 
 
We demonstrate that dipolar magnetic interactions inhibit boundary cluster formation, thereby increasing the critical packing fraction and leading to the emergence of new collective behaviors.

\begin{figure}[t!]
    \centering
    \includegraphics[width=1\linewidth]{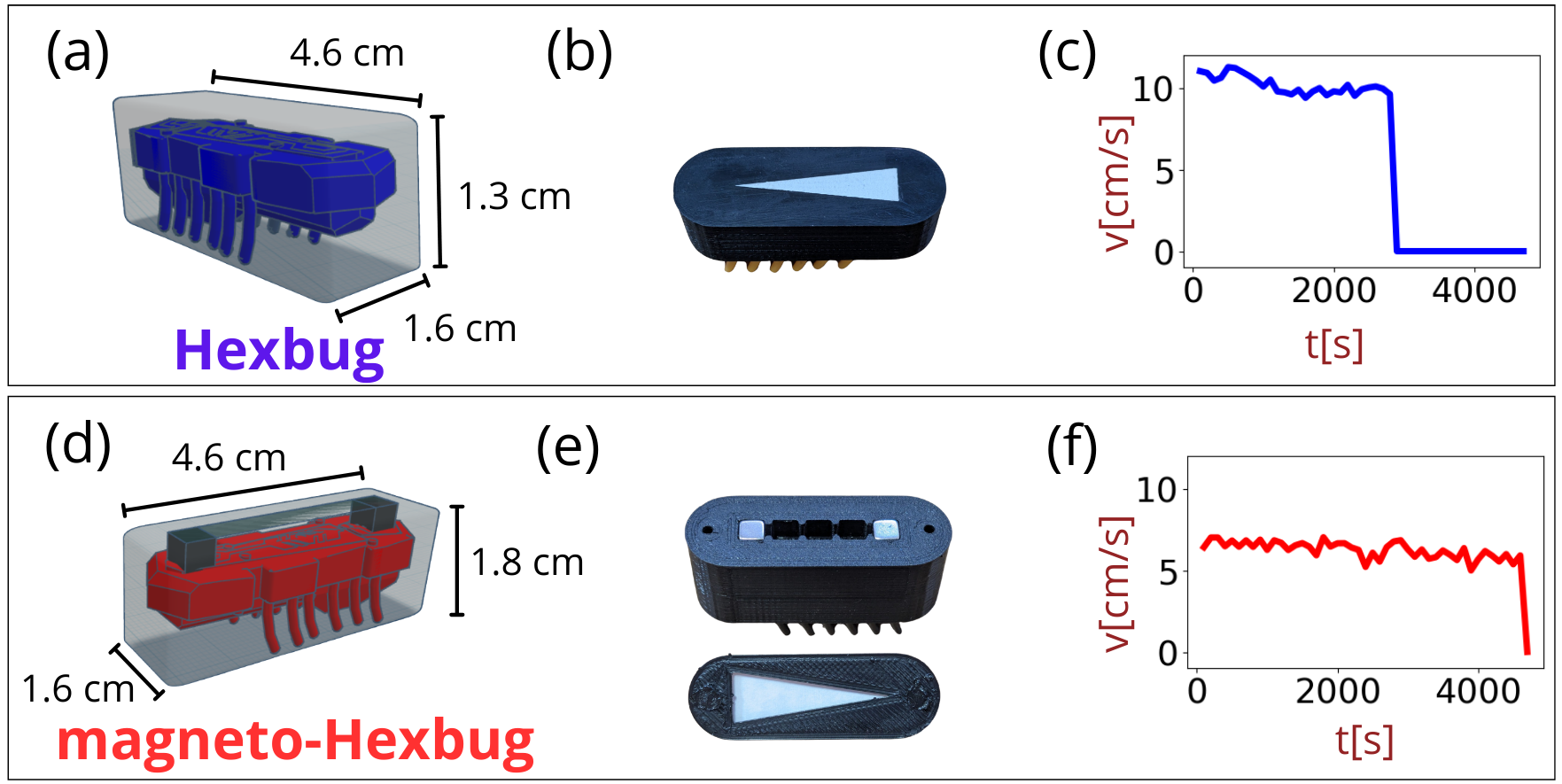}
    \caption{Experimental setup: (a,d) 3D models of the Hexbugs and the magneto-Hexbug. The shell of the magnetic bug is taller to accommodate two or more small neodymium magnets. (b,e) Images of the Hexbug and the magneto-Hexbug showing their shells and the white marker used in the tracking process. (c,f) Temporal profiles of the spped for magnetic and non-magnetic particles.}
    \label{fig:setup}
\end{figure}

{\em Experiments}.
We performed experiments with two types of elongated ``particles'', magnetic and non-magnetic Hexbugs (see Fig. \ref{fig:setup}).
A 3D-printed shell on both units ensures smoother contact interactions.
Magneto-bugs are equipped with two small permanent neodymium magnets (cubic shape, 5~mm wide, with a magnetic moment of 0.09~A\,m$^2$) 
placed with their polar vector pointing in the same direction as the main axis of the particle, one close to the head and one  to the tail.
Slight differences in shell weight result in variations in the characteristic speeds and lifetimes of the two types
(Fig. \ref{fig:setup}).
The experiments involve tracking the dynamics of $N$ (up to $39$) Hexbugs within a circular arena 
with diameter $30$ cm, 3D-printed in Polylactic Acid (PLA), with boundaries and no roof. 
The floor is smoothed with an adhesive Polyvinyl Chloride (PVC) film.

\begin{figure*}[ht]
    \centering
    \includegraphics[width=1\linewidth]{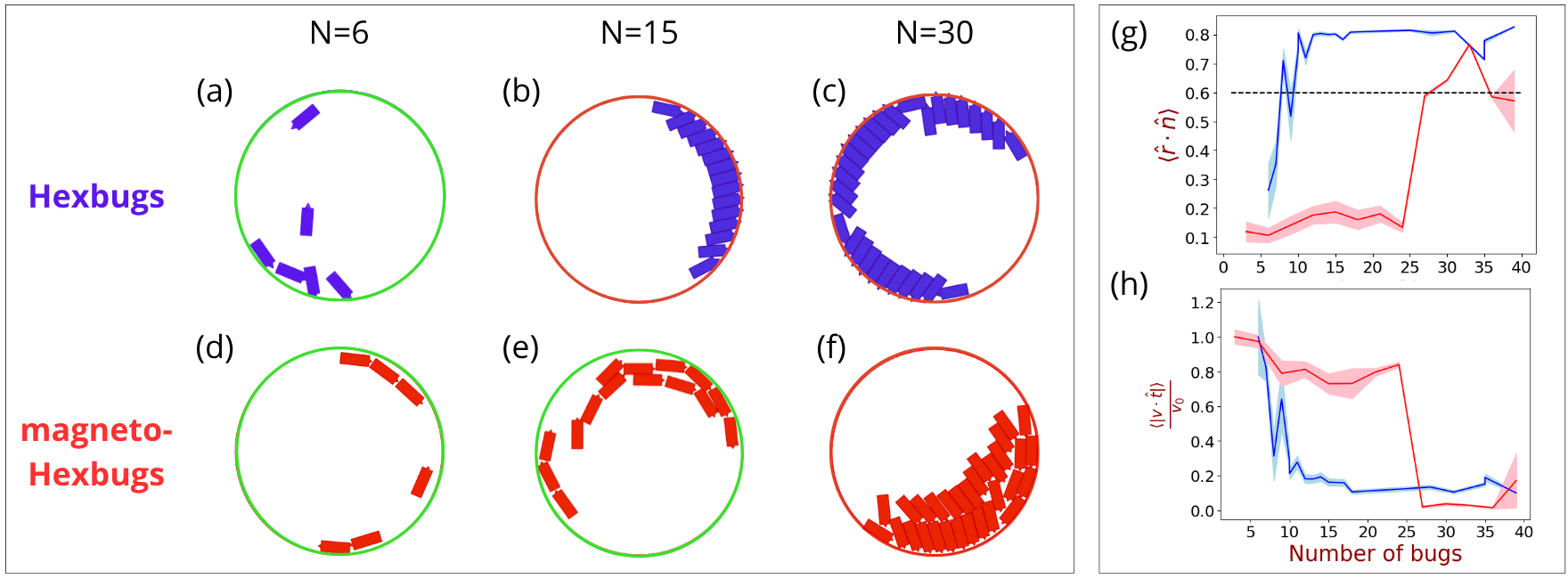}
    \caption{Experimental results: (a–c) Experimental frames showing the final configurations for $N = 6, 15,$ and $30$ Hexbugs. (d–f) Experimental frames showing the final configurations for $N = 6, 15,$ and $30$ magneto-Hexbugs. (g) Plot of $\langle \hat{r} \cdot \hat{n} \rangle$, where $\hat{r}$ is the unit vector pointing from the arena center to the particle’s center of mass, and $\hat{n}$ is the unit vector of the particle’s orientation. (h) Plot of $\langle |v \cdot \hat{t}| \rangle / v_0$, where $\hat{t}$ is the unit vector perpendicular to $\hat{r}$ (i.e., locally tangent to the boundary), $v$ is the particle velocity and $v_0$ is the mean velocity measured at the lowest density. Blue curve: experiments with Hexbugs; red curve: experiments with magneto-Hexbugs. Green and red boundaries in (a-f) are used to highlight the final configuration, without or with boundary-clusters respectively. } 
    \label{fig:resultsexp}
\end{figure*}

The main experimental results are recapitulated in Fig.~\ref{fig:resultsexp}. Following the terminology of~\cite{deblais2018boundaries}, we observe "boundary polar clusters", in both magnetic and non-magnetic cases, provided that the number of particles is large enough, see Fig.~\ref{fig:resultsexp} (b,c,f). These clusters are characterized by a majority of particles with aligned polar vectors - pointing roughly perpendicular to the boundary and towards the outside of the arena - and close-packed positions along the boundary. All clusters display two "confining" particles at their extremes which push inward, providing stability of the cluster. Variations of this typical arrangement include the presence of defects (see Fig.~\ref{fig:resultsexp}c) as well as multilayering (see Fig.~\ref{fig:resultsexp}f). For small numbers of particles, the clusters do not form, and the particles are typically found moving along trajectories which are (roughly) parallel to the boundary: in the case of non-magnetic particles this "gas"-phase is without an evident positional order, while in the magnetic case the particles typically display "trains" of variable length, not far from the boundary. 
We observe that, in the low-density regime, boundary-following trajectories tend to exhibit {\it anti-chirality}, moving in the direction opposite to the intrinsic chirality of unconfined Hexbug particles. This occurs because chiral motion more effectively stabilizes boundary trajectories with opposite chirality.

\begin{figure}[hb]
    \centering
    \includegraphics[width=0.7\linewidth]{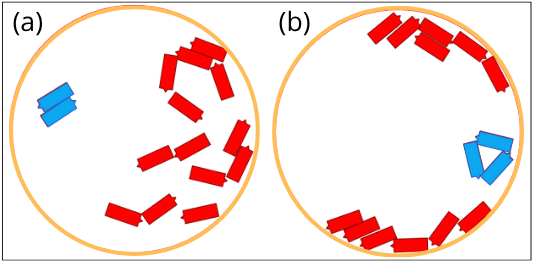}
    \caption{
    Typical rotating structures (blue particles) observed in experiments resulting from dipole-dipole magnetic interactions: (a) a rotating pair and (b) a three-particle vortex.
    }
    \label{fig:rot_structures}
\end{figure}

The main difference between the non-magnetic and magnetic case, which represents our main novel experimental observation, is the increase of the clustering threshold density for the magnetic particles. This result 
is made clear in Figs.~\ref{fig:resultsexp}g,h, where two different order parameters are shown as a function of the number of particles, for the non-magnetic (blue data) and magnetic (red data) particles. In Fig.~\ref{fig:resultsexp}g we show $\langle \hat{r} \cdot \hat{n} \rangle$, where $\hat{r}$ is the unit vector of the center of mass position of each particle, assuming the center of the arena as the origin, while $\hat{n}$ is the unit vector representing the polar orientation of each particle. The scalar product is first averaged over all particles at a given time, and then averaged over a long realization of the experiment (corresponding to about two-thirds of the total duration, i.e., approximately 10 minutes). 

The figure shows a clear transition in the order parameter from a majority of particles aligned to the boundary 
($\hat{r} \cdot \hat{n} \approx 0.1$) to a majority of particles perpendicular to it ($\hat{r} \cdot \hat{n} \gtrsim  0.6$), signaling the emergence of polar boundary clusters. The transition occurs  at $N = N^*_{nm} \approx 10$ for non-magnetic particles and at $N = N^*_{m} \approx 27$ for magnetic ones.

The other order parameter we have analyzed is $\langle |v \cdot \hat{t}| \rangle/v_0$, where $\hat{t}$ is the unit vector perpendicular to $\hat{r}$, i.e. parallel to the boundary in the proximity of the particle, and $v$ is the velocity vector of the particle (where $v_0$ denotes the mean velocity observed at the lowest density). This order parameter signals the same transition - from $\sim 1$ (particles moving parallel to the boundary) to $\sim 0$ (particles not moving or moving perpendicular to it) - at the same critical sizes $N^*_{nm} <  N^*_{m}$. This order parameter is related to velocity instead of orientation: it may appear that the two are usually parallel, but this is not true at high density, e.g. in the cluster phase, and in fact it is seen that a residual velocity parallel to the boundary is observed. In fact, the clusters crawl along the boundary thanks to a small asymmetry in their internal structure, at a speed which is significantly lower than the free speed of particles.
We also note the emergence of several characteristic transient structures during the dynamic evolution of magnetic Hexbugs, such as rotating pairs, linear trains, and vortices composed of three or more particles (see Fig.~\ref{fig:rot_structures}), never observed in non-magnetic bugs.

\begin{figure*}[ht]
    \centering
    \includegraphics[width=1\linewidth]{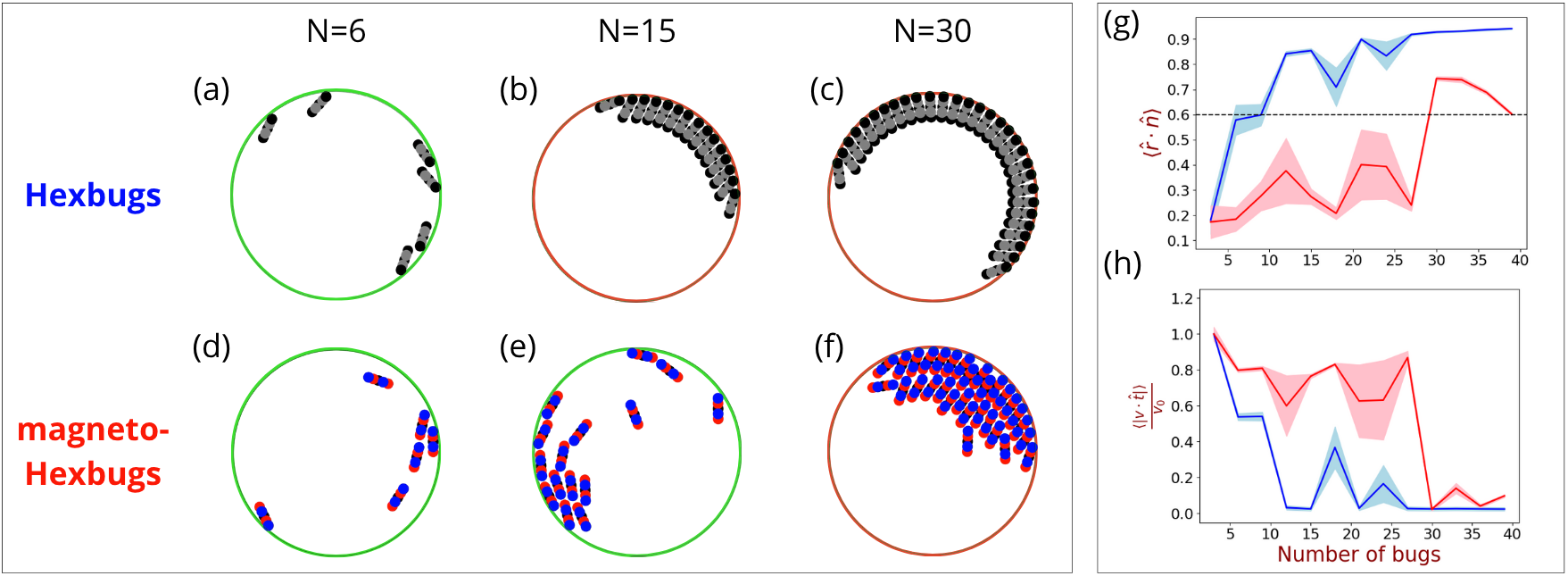}
    \caption{Simulation results: (a–c) Simulation frames showing the final configurations for $N = 6, 15$ and $30$ particles. (d–f) Simulation frames showing the final configurations for $N = 6, 15$ and $30$ magnetic particles. In all cases (a-f), each hexbug is represented by its $5$ centers of force, depicted as coloured discs (gray and black for the non-magnetic, red and blue for the magnetic ones); the head of a hexbug is the only disc which is fully visible (black for the non-magnetic, blue for the magnetic ones). (g) Plot of $\langle \hat{r} \cdot \hat{n} \rangle$, where $\hat{r}$ is the unit vector pointing from the arena center to the particle’s center of mass, and $\hat{n}$ is the unit vector of the particle’s orientation. (h) Plot of $\langle |v \cdot \hat{t}| \rangle / v_0$, where $\hat{t}$ is the unit vector perpendicular to $\hat{r}$ (i.e., locally tangent to the boundary), $v$ is the particle velocity and $v_0$ is the mean velocity measured at the lowest density. Blue curve: simulations with non-magnetic particles; red curve: simulations with magnetic particles.}
    \label{fig:resultssim}
\end{figure*}


{\em Simulations}.
Numerical simulations are performed considering $N$ active Brownian particles with elongated shape and confined in a 2D circular domain.
Each particle $i$ is described by its center of mass position ${\bf r}_i$,  orientation $\hat{\mathbf{n}}_i=(\cos(\varphi_i),\sin(\varphi_i))$ and by $M\!=\!5$ centers of force along its principal axis, whose positions are   ${\bf r}_i^\alpha$ ($\alpha=1,\dots,M$).
 
Magnetic particles are equipped with two magnets with 
magnetic moment ${\mathbf{m}}_i=m \hat{\mathbf u}_i$, whose orientation
coincides with the particle's orientation $\hat{\mathbf u}_i=\hat{\mathbf n}_i$ (but in general could be different), and position ${\bf r}_i^p$ ($p=1,2$).
The equations of motion of the $i$-th particle in the overdamped regime are
\begin{align}
\gamma \dot{\mathbf{r}}_i &= -\nabla_i U + \mathbf{F}_{{i,active}} \\[6pt]
\gamma_r \dot{\varphi}_i &= 
\mathbf{T}_i \cdot \hat{\mathbf{z}}
+ \beta \, (\hat{\mathbf{n}}_i \times \dot{\mathbf{r}}_i) \cdot \hat{\mathbf{z}} 
+ \gamma_r \eta
\end{align}
where $\gamma$ and $\gamma_r$ are the translational and rotational friction coefficients, 
$\mathbf{F}_{i,active}=\gamma v_0 \hat{\mathbf{n}}_i$ is the self-propulsion force of each individual unit 
($v_0$ is the self-propelled speed),
$\beta$ is the strength of the self-alignment term~\cite{baconnier2025self}, 
and $\eta$ is a Gaussian white noise with zero mean $\langle \eta(t)\rangle=0$ and variance 
$\langle \eta(t)\eta(s)\rangle =2D_r\delta(t-s)$, with $D_r$ the rotational diffusion constant.
The potential energy 
$U = \sum_{i < j} \left( \sum_{\alpha,\beta} U_{\text{S}}({r}_{ij}^{\alpha \beta}) + \sum_{p,q} U_{\text{M}}(\mathbf{r}_{ij}^{pq},\hat{\mathbf u}_i,\hat{\mathbf u}_j)\right)$ accounts for both steric interactions  and  dipole–dipole interactions.
In the previous expression the sum on $\alpha,\beta$ runs over centers of force  and  the sum on $p,q$  over magnets of each particle,  
$\mathbf{r}_{ij}^{\alpha \beta}=\mathbf{r}_i^\alpha-\mathbf{r}_j^\beta$,
$r_{ij}^{\alpha \beta}=\lvert \mathbf{r}_{ij}^{\alpha \beta}\rvert$ 
and similarly  for $\mathbf{r}_{ij}^{p q}$.
The steric term is a truncated repulsive soft potential, 
$U_{\text{S}}(r) = 4 \varepsilon (\sigma/r)^{12}\ \theta(r_c-  r)$,
where $(\epsilon, \sigma)$ set the energy and length scales of the interaction, $r_c$ is a cut-off radius and $\theta$ is the Haviside step function.
The magnetic interactions are described by the dipole-dipole energy
\begin{equation}
U_{\text{M}}(\mathbf{r},\hat{\mathbf{u}}_1,\hat{\mathbf{u}}_2) = 
\frac{\mu_0 m^2}{4\pi r^3} \left[ \hat{\mathbf{u}}_1 \cdot \hat{\mathbf{u}}_2
- 3 \frac{(\hat{\mathbf{u}}_1 \cdot \mathbf{r}) (\hat{\mathbf{u}}_2 \cdot \mathbf{r})}{r^2} \right] .
\end{equation}
The boundary-particle interaction is modeled using the image particle method 
\cite{Angelani_2009,Paoluzzi_2015,Paoluzzi_2020}.
$\mathbf{T}_i$ represents the torque that arises from both steric and dipole–dipole interactions 
(see Supplementary Material for details). We set the model parameters to describe the experimental conditions. In particular, we choose them such that
the P\'eclet number, $\mathrm{Pe} = v_0 / (D_r \sigma)$, takes the value $\mathrm{Pe} = 70$ for non-magnetic Hexbugs and 
$\mathrm{Pe} = 50$ for magnetic ones, the dimensionless strength of the magnetic interaction is
$\Lambda = \mu_0 m^2 / (4 \pi \, \sigma^3 \, D_r \gamma) = 4.7$, 
the reduced self-alignment strength is $B = \beta \sigma / \gamma_r = 0.35$, the friction ratio
is $\gamma/\gamma_r=2/3$.

The results of the simulations are shown in Fig.~\ref{fig:resultssim}, which represents the numerical counterpart of the experimental results of Fig.~\ref{fig:resultsexp}.
We first notice that the observed configurations are quite similar to the experimental ones, with the exception of a lack of defects in the boundary clusters of the non-magnetic case. All other observations, such as trains in the low density magnetic system, single layers of boundary clusters in the non-magnetic case opposed to bilayers in the magnetic one, are faithfully reproduced. 

\begin{figure}[hbt!]
    \centering
    \includegraphics[width=\linewidth]{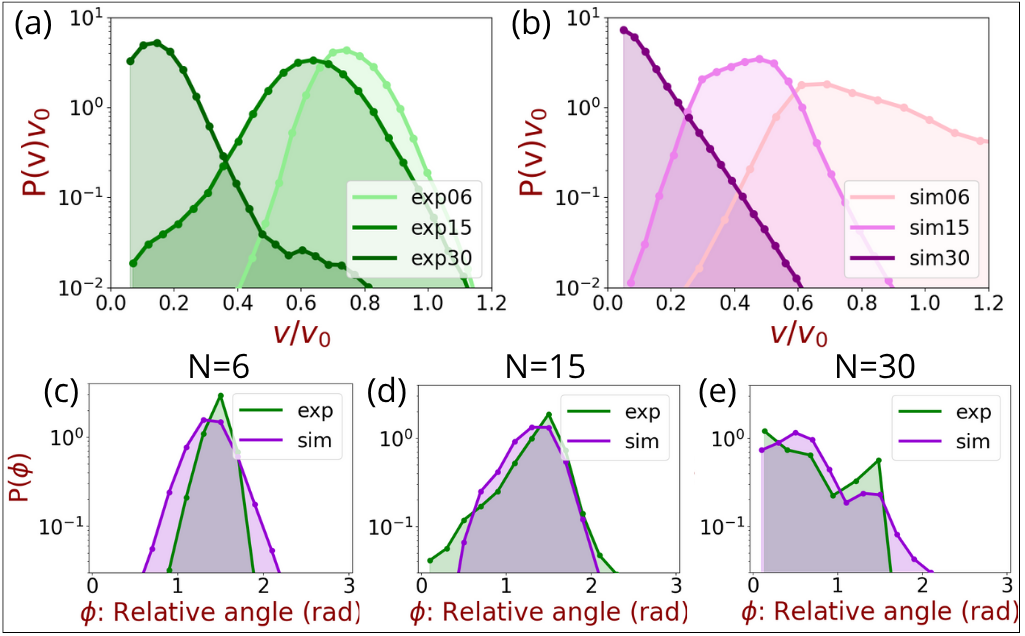}
    \caption{Comparison between experiments and simulations for magneto-Hexbugs. (a,b) Velocity distributions for $N = 6, 15,$ and $30$ bugs in experiments (a) and simulations (b). Velocities are normalized by $v_0$, defined as the typical body length of a Hexbug divided by the characteristic time it takes to travel this distance. (c–e) Probability distributions of the relative angle $\gamma$ (modulo $\pi$), shown in green for experiments and violet for simulations. The angle $\gamma$ is defined as the difference between the particle orientation in the lab frame and the angle of the vector from the arena center to the particle’s center of mass. The cases with N = 6, 15 and 30 are shown. }
    \label{fig:sim_exp_comp}
\end{figure}

The behavior with $N$ of the two aforementioned order parameters is also in fair agreement with the experiments, with transitions from non-clustered to clustered phases at $N^*_{nm}<N^*_{m}$ with values compatible with the experiments. We note that in the simulation the transition of the non-magnetic system is less sharp, i.e. the growth of the order parameter is smoother and, in particular for $\langle \hat{r}\cdot \hat{n}\rangle$, it does not immediately saturates for values $N \gtrsim N^*_{nm}$.\\
\\
To enable a more detailed comparison between experiments and simulations in the magnetic case, Fig.~\ref{fig:sim_exp_comp} presents 
the velocity distributions (a,b) as well as the distributions of the relative angle between particle orientation and position vector,
$\phi=\cos^{-1} (\hat{\mathbf n}\cdot \hat{\mathbf r})$, for 6, 15, and 30 bugs (c–e). 
Panels (a) and (b) show that, in both simulations and experiments, increasing system density leads to a shift in the velocity distribution toward lower values, consistent with the formation of particle clusters.
Panels (c–e) demonstrate that magneto-Hexbugs  at low densities predominantly move parallel to the boundary (indicated by a peak near $\pi/2$). As N increases, the emergence of clusters with radially oriented particles results in angle distributions shifting toward zero.
The small peak near $\pi/2$ observed at $N=30$ indicates that some particles within the cluster maintain alignment with the boundary.

{\em Conclusions}.
We investigated magnetic active systems by conducting experiments with modified Hexbug toy robots, each equipped with a shell containing small permanent magnets. 
We discovered that magnetic interactions produce a {\it fluidization} effect on the agglomerates that usually form near boundaries in standard active systems. 
More specifically, we  found that boundary clusters form at higher densities with respect to the non-magnetic case. 
Furthermore, magnetic interactions produce a wide variety of particular transient configurations, such as train-like structures, rotating pairs and vortices.
We confirmed these results through numerical simulations, implementing a model of active magnetic particles that reproduced the experimental conditions, finding very good agreement between them.

An intriguing question is whether this {\it fluidization} effect persists in bulk systems, potentially influencing the MIPS phenomenon~\cite{spera2024nematic}.
Furthermore, our design enables the systematic study of various magnetic particle configurations, allowing for controlled modifications in the position, number, and orientation of the magnets. This flexibility opens the door to exploring a broad range of interaction scenarios and collective behaviors in magnetic active matter.

\begin{acknowledgments}
The authors acknowledge discussions with Lorenzo Caprini.   A.G., M.P. and A.P. acknowledge funding from the Italian Ministero dell’Universit\`a
e della Ricerca under the programme PRIN 2022 (``re-ranking of the final lists''), number 2022KWTEB7, cup. B53C24006470006.
L.A. acknowledges funding from the Italian Ministero dell’Università e della Ricerca under the programme
PRIN 2020, number 2020PFCXPE.
\end{acknowledgments}

\bibliography{biblio}


\onecolumngrid
\section*{Supplementary Material}

\subsection{Experimental setup}
The Hexbugs we used are described in Figure~\ref{fig:setup}. 
The shells are slightly different for magnetic and non-magnetic particles, and this affects the typical speed and battery lifetime of the two species: non-magnetic ones are lighter and therefore can express a higher speed ($\sim 10$ cm/s) but a shorter lifetime (less than a hour); magnetic bugs are heavier, resulting in a smaller speed ($\sim 6$ cm/s) with a longer lifetime (almost $1.5$ hours). The relation between weight, speed and lifetime is not simple to rationalize since it involves friction which can be both beneficial and detrimental for bugs' propulsion and energetic efficiency (see for instance~\cite{koumakis2016mechanism}). We note that, while the battery lifetime is consistent with the observations reported in~\cite{deblais2018boundaries} (their Fig.~S1), the speed of our Hexbugs is generally lower but more stable over time compared to that study. This discrepancy can be attributed to the specific experimental conditions under which the results in Figure~\ref{fig:setup} were obtained: both magnetic and non-magnetic Hexbugs were confined to a circular track to facilitate motion tracking. Nevertheless, this setup is useful to quantify, under identical conditions, the velocity ratio between the two types of particles.

We have detected, as in all previous studies with hexbugs~\cite{deblais2018boundaries,van2023self} an evident tendency of the bugs to follow curve paths with each bug exhibiting a systematic curvature (in both radius and orientation) with fluctuations, not easy to be fully characterised in a finite size arena as ours. A population of bugs seems to contain a distribution of curvatures which is roughly symmetric around zero. We adopt for this curvature the term ``chirality'', in accordance with~\cite{deblais2018boundaries}.

We also directly measured self alignment ~\cite{baconnier2025self, musacchio2025self, musacchio2025flocking} and incorporated it into the model to capture the full phenomenology.

Our bugs have a non negligible inertia, for instance some bounce-back can be observed in collisions with the wall or in direct collisions between two Hexbugs. However the simulations we have performed, neglecting inertia, are able to provide us with a fair qualitative account of our main experimental observations, suggesting that it is not a crucial ingredient of the phenomenology. We underline that we observe a transition from a disordered "gas-like" phase to a boundary-clustered phase, without coexistence. This is coherent with the claim, in~\cite{deblais2018boundaries}, that phase coexistence requires inertia and that in the overdamped limit their simulations display pure clusters at the boundaries (their Fig. S7).

In the non-magnetic case, all hexbugs were placed inside a tray that tightly fits $3$ rows of $13$ bugs; after all bugs are switched on, the tray is lifted; the bugs in one row are all oriented towards another row, causing them to immediately break up any order and diffuse into the arena. A different strategy was adopted in the magnetic experiments: due to magnetic interactions, placing the bugs in a compact initial configuration would have influenced the final state of the system. In this case, the magneto-Hexbugs were allowed to enter the arena one by one through an external railway, and the experiment effectively started once all the selected bugs were inside the arena. Moreover, to ensure that the experiments were not affected by a drop in the battery of some Hexbugs, we kept track of each bug’s hours of activity and replaced the batteries so as to avoid observing a sudden decrease in the particles’ speed.

\subsection{Simulation model}
We have tuned the parameters of the model in order to reproduce the behavior observed in the experiments, in particular the spatial structures and the order parameters as function of $N$. Even if the number of parameters is large, our requirement of obtaining the full phenomenology with two main phases at different values of $N$ put a severe constraint. 
The forces shaping the translational dynamics of the particles are derived from the steric and magnetic potentials introduced in the \textit{Simulation} section. Moreover, in order to correctly compute the torque $T_i$ appearing in Eq.~(2), we have considered three different contributions. The first one arises from steric interactions, which are computed for all the $M$ centers of force along the principal axes of each particle and subsequently added to the evolution of the particle’s orientation. The other contributions originate from magnetic interactions. 
Considering two magnetic dipoles belonging to different particles located at positions $\mathbf{r}_i^{p}$ and $\mathbf{r}_j^{q}$, 
the magnetic field produced by magnet $q$ of particle $j$ on the magnet $p$ of particle $i$ is 
$\mathbf{B}(\mathbf{r}_{ij}^{pq},\hat{\mathbf{u}}_j)$, where
$\mathbf{r}_{ij}^{pq}=\mathbf{r}_i^{p}-\mathbf{r}_j^{q}$, $\hat{\mathbf{u}}_j$ is the orientation of the magnet $q$ of particle $j$ and
\begin{equation}
\mathbf{B}(\mathbf{r},   \hat{\mathbf{u}}) = \frac{\mu m}{4 \pi r^3} \left[ \frac{3 \mathbf{r} (\mathbf{r}\cdot \hat{\mathbf{u}})}{r^2} - \hat{\mathbf{u}}\right] .
\end{equation}
The torque exerted on particle $i$ by the magnets of particle $j$ can be expressed as
\begin{equation}
    \mathbf{T}_{ij} = \sum_{p,q} m\hat{\mathbf{u}}_i \times \mathbf{B}(\mathbf{r}_{ij}^{pq},\hat{\mathbf{u}}_j) .
\end{equation}
The second contribution to the torque due to magnetic interactions arises from the fact that the magnetic forces governing the translational dynamics described in Eq.~(1) are applied not at the particle’s center of mass, but at the center of the magnet.

To test cluster stability, the system was initialized in a pre-assembled configuration and then allowed to evolve as already done in~\cite{deblais2018boundaries}. At low densities, such aggregates are unstable: after some time, particles move apart and begin circular motion along the boundaries. At high densities, instead, the clusters remain intact. The morphology also differs between magnetic and non-magnetic cases: without magnetic interactions, the aggregate forms a single layer, as in the experiments, whereas in the magnetic case a second layer develops once enough magneto-Hexbugs occupy the first one, again consistent with the observations.
 
We managed to obtain a good agreement even without chirality (we mention the fact that self-alignment was not considered in~\cite{deblais2018boundaries} and in the same work chirality was observed to inhibit clustering and polar ordering). 
On the contrary we have realized that a crucial ingredient is the distribution of particle speeds, 
otherwise certain structure (such as boundary-running trains) are too stable and never break in favor of clusters. The speed distribution that we found to be optimal for comparison with experiments is a Gaussian distribution with two thresholds, in order to avoid having bugs that are too slow or too fast. The mean value of the Gaussian distribution is 0.7 for the non magnetic-Hexbugs and 0.5 for the magnetic ones, for which the resulting ratio is consistent with the experimental data reported in Figure~\ref{fig:setup}.

\end{document}